\documentclass[twocolumn]{aastex631}

\usepackage{color}
\usepackage{url}
\usepackage{hyperref}
\usepackage{xspace}
\usepackage{soul}
\usepackage{hyperref}

\newcommand{\source}{{Swift~J1727.8$–$1613}\xspace}

\newcommand{\referee}[1]{\textcolor{black}{#1}}

\shortauthors{Ingram et al.}

\begin{document}

\title{Tracking the X-ray Polarization of the Black Hole Transient Swift~J1727.8$–$1613 during a State Transition}

%
\author[0000-0002-5311-9078]{Adam Ingram}
\affiliation{School of Mathematics, Statistics, and Physics, Newcastle University, Newcastle upon Tyne NE1 7RU, UK}
\author[0009-0005-6609-5852]{Niek Bollemeijer}
\affiliation{Anton Pannekoek Institute for Astronomy, Amsterdam, Science Park 904, NL-1098 NH, The Netherlands}
\author[0000-0002-5767-7253]{Alexandra Veledina}
\affiliation{Department of Physics and Astronomy, 20014 University of Turku, Finland}
\affiliation{Nordita, KTH Royal Institute of Technology and Stockholm University, Hannes Alfv\'ens v\"ag 12, SE-10691 Stockholm, Sweden}
\author[0000-0003-0079-1239]{Michal Dov\v{c}iak}
\affiliation{Astronomical Institute of the Czech Academy of Sciences, Bo\v{c}n\'{i} II 1401/1, 14100 Praha 4, Czech Republic}
\author[0000-0002-0983-0049]{Juri Poutanen}
\affiliation{Department of Physics and Astronomy,  20014 University of Turku, Finland}
\author[0000-0002-1532-4142]{Elise Egron} 
\affiliation{INAF Osservatorio Astronomico di Cagliari, Via della Scienza 5, 09047 Selargius (CA), Italy}
\author[0000-0002-7930-2276]{Thomas D. Russell} 
\affiliation{INAF, Istituto di Astrofisica Spaziale e Fisica Cosmica, Via U. La Malfa 153, I-90146 Palermo, Italy}
\author[0000-0002-7586-5856]{Sergei A. Trushkin} 
\affiliation{Special Astrophysical Observatory of the Russian Academy of Sciences, Nizhnij Arkhyz, 369167, Karachayevo-Cherkessia, Russia}
\author[0000-0002-6548-5622]{Michela Negro} 
\affiliation{Department of Physics and Astronomy, Louisiana State University, Baton Rouge, LA 70803, USA}
\author[0000-0003-0411-4243]{Ajay Ratheesh}
\affiliation{INAF Istituto di Astrofisica e Planetologia Spaziali, Via del Fosso del Cavaliere 100, 00133 Roma, Italy}
% ...

\collaboration{123}{(IXPE Collaboration)} 


\begin{abstract}
We report on a\referee{n observational} campaign on the bright black hole X-ray binary \source centered around five observations by the Imaging X-ray Polarimetry Explorer (IXPE). \referee{These observations track for the first time} the evolution of the X-ray polarization of a black hole X-ray binary across a hard to soft state transition. The 2--8 keV polarization \referee{degree} decreased from $\sim$4\% to $\sim$3\% across the five observations, but \referee{the polarization angle remained oriented} in the North-South direction throughout. \referee{Based on observations with} the Australia Telescope Compact Array (ATCA), we \referee{find that} the intrinsic 7.25 GHz radio polarization align\referee{s with the X-ray polarization}. Assuming the radio polarization aligns with the jet direction (which can be tested in the future with \referee{higher spatial resolution} images \referee{of the jet}), \referee{our results imply} that the X-ray corona is extended in the disk plane, rather than along the jet axis, for the entire hard intermediate state. This in turn implies that the long ($\gtrsim$10 ms) soft lags that we measure with the Neutron star Interior Composition ExploreR (NICER) are dominated by processes other than pure light-crossing delays. Moreover, we find that the evolution of the soft lag amplitude with spectral state \referee{does not follow the} trend seen for other sources, implying that \source is a member of a hitherto under-sampled sub-population.
\end{abstract}



\section{Introduction}
\label{sec:intro}

Black hole (BH) X-ray binaries display transitions in their X-ray spectra on timescales of days to months. The \referee{energy spectrum in the} \textit{soft state} is dominated by a multi-temperature blackbody component originating from an optically thick, geometrically thin accretion disk \citep{Shakura1973,Novikov1973}. \referee{In contrast, the emission in the} \textit{hard state} is dominated by a hard power law with high-energy cut off, originating from Compton up-scattering of disk seed photons in a cloud of hot electrons located close to the BH \citep{Thorne1975,Sunyaev1979}, commonly referred to as the \textit{X-ray corona}. The \referee{origin, geometry, location, and physical properties of the coronal gas responsible for the power-law X-ray emission} is still an intensely debated topic \citep[\referee{e.g.,}][]{Poutanen2018,Bambi2021}. It has\referee{, for example,} been suggested to be a (patchy) layer located above the disk (the \textit{sandwich model}, \citealt{Galeev1979,Haardt1993,Stern1995}), a large scale height accretion flow inside a truncated disk (the \textit{truncated disk model}, \citealt{Eardley1975,Esin1997,Poutanen1997}) or the (vertically extended or compact) base of the jet \citep{Miyamoto1991,Martocchia1996,Markoff2005}. Characteristic reflection features including an iron K$_\alpha$ line at $\sim$6.4~keV and a broad Compton hump at $\sim$20--30~keV are also observed \citep{Miller2007}, which result from coronal X-rays irradiating the disk and being reprocessed and re-emitted into the observer's line of sight \citep[\referee{e.g.,}][]{Matt1991,Garcia2010}. Although the iron line is narrow in the local restframe, rapid orbital motion and the gravitational pull of the BH distort its shape \citep{Fabian1989}.

\begin{figure*} 
\centering
\includegraphics[width=0.92\linewidth]{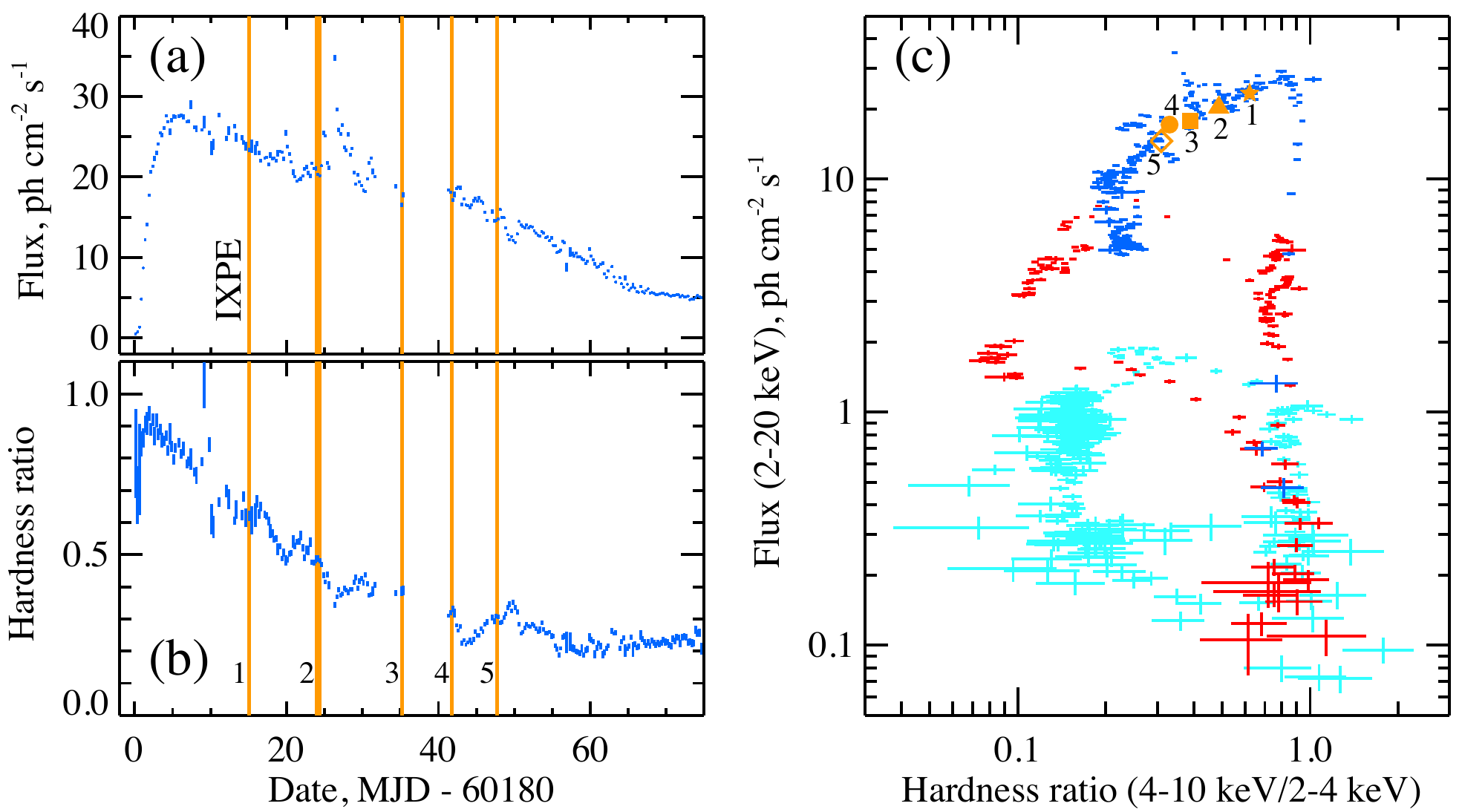}
\caption{Evolution of X-ray properties of \source during the outburst, determined with MAXI \citep{Matsuoka2009}. 
Vertical orange strips show the times of IXPE observations. 
(a) MAXI light curve in the 2--20 keV range.  
(b) Hardness ratio of the photon flux in the 4--10 keV band to that in the 2--4 keV band. 
(c) Hardness-intensity diagram. 
Blue crosses show the evolution of \source during the current outburst. 
The positions of the source on the diagram during \referee{the} five IXPE observations \referee{(as numbered) are shown by the} orange symbols. For comparison, we also plot the time evolution of MAXI~J1820+070 during its 2018 outburst with red crosses and GX~339$-$4 during its 2010 outburst with cyan crosses. This plot is designed as an update of Figure 1 in \citet{Veledina2023a}.} \label{fig:maxi}
\end{figure*}

Most BH X-ray binaries spend the majority of their time in quiescence and occasionally undergo outbursts of dramatically increased brightness \citep{Done2007,Tetarenko2016}. The progress of an outburst can be tracked with a hardness intensity diagram (HID), whereby hardness (the ratio of X-ray counts in \referee{a} harder band to the counts in \referee{a} softer band) is on the x-axis and X-ray flux is on the y-axis \citep[\referee{e.g.,}][]{Homan2001,Fender2004}. An archetypal outburst traces out a q-shape on the HID, moving in an anti-clockwise direction (see Figure~\ref{fig:maxi}c for some examples) as it rises from quiescence, then transitions from hard to soft state via the \textit{intermediate state} and eventually back through the intermediate state to the hard state and finally quiescence \citep[\referee{e.g.,}][]{Homan2005,Belloni2010}.

The X-ray flux displays rapid variability with properties tightly correlated with the spectral changes \citep{VDK2006}. In the hard and intermediate states, quasi periodic oscillations (QPOs) are often observed, which appear in the power spectrum as narrow, harmonically related peaks (see \citealt{Ingram2019b} for a review). Low frequency QPOs, with frequencies in the range $\sim$0.01--20~Hz, are classified into three categories \citep{Casella2005}. Type-C QPOs are by far the strongest and most commonly observed. They first appear in the hard state superimposed on top of band limited noise that features a roughly constant variability amplitude between low and high frequency breaks. Type-B QPOs are marked by the almost complete disappearance of the band limited noise. The presence of either Type-C or Type-B QPOs is used to sub-classify the intermediate state into the hard intermediate state (HIMS) and soft intermediate state (SIMS), \referee{respectively}. The soft state features a very low level of variability, except for occasional QPOs including Type-A QPOs, which are fairly broad and weak. Whereas the spectral evolution tracked by the HID can vary from source to source, or even from outburst to outburst of a given source \citep[e.g.,][]{Dunn2010}, the evolution of the band limited noise appears to be universal \referee{\citep[e.g.,][]{Belloni2010}} and can be tracked with just one parameter \referee{---} the power spectral hue \citep{Heil2015a}. \referee{The hue} describes the relative amplitude of the power spectrum in four broad `power-color' frequency bands (see Section \ref{sec:hue} for a detailed explanation).



Time lags between variations of the flux in different energy bands can be measured using the cross spectrum \citep{vanderKlis1987}. The time lags measured from the cross spectrum are a function of Fourier frequency, enabling lags between variations on different characteristic variability timescales to be separated out. At low Fourier frequencies (long variability timescales), variations in hard X-rays are observed to lag those in soft X-rays (so-called hard lags; \referee{e.g.,} \citealt{Miyamoto1988,Nowak1999}). This \referee{lag} is
commonly attributed \referee{to the} inward propagation of accretion rate fluctuations \citep[\referee{e.g.,}][]{Lyubarskii1997,Kotov2001,Arevalo2006,Ingram2013}, but has been alternatively interpreted as light crossing lags between successive Compton scattering orders in a highly extended corona \citep[e.g.,][]{Kazanas1997,Reig2015} or as spectral evolution during X-ray flares \citep{Poutanen1999,Kording2004}. At high Fourier frequencies ($\nu\gtrsim 1-10$ Hz, short variability timescales), soft lags are instead observed \citep[\referee{e.g.,}][]{Uttley2011,DeMarco2015,DeMarco2017}. The soft lags are typically attributed to \textit{reverberation} \citep{Reynolds1999,Poutanen2002,Uttley2014}, in that reflected photons take a longer path to the observer than \referee{coronal} photons \referee{that directly reach the observer}, and the reflection spectrum includes a soft excess. The soft lags follow a common evolution with power spectral hue across the population of BH X-ray binaries that sees them sharply increase during the HIMS from as low as $\sim 0.5$~ms to as high as $100$~ms \citep{Wang2022}. Under the reverberation interpretation, this would require the vertical extent of the corona to dramatically increase during the HIMS \citep{deMarco2021}, which \citet{Wang2021} suggested may be causally connected to the discrete jet ejections observed at radio wavelengths days later. However, other processes may contribute to, and may even dominate, the soft lag signal, including a complex interplay of broadband spectral components \citep{Veledina2018,Kawamura2023}. Further information is therefore required to determine the relative importance of light-crossing delays and other processes.


The Imaging X-ray Polarimetry Explorer \citep[IXPE;][]{Weisskopf2022} now enables sensitive X-ray polarimetry, providing a novel diagnostic of the coronal geometry and its evolution with spectral state. In energy bands in which we see photons after multiple Compton up-scatterings, the net polarization of the optically-thin corona aligns with its minor axis, such that, for example, a horizontally extended corona will be vertically polarized \citep{Poutanen1996,Ursini2022}. 
This is because photons propagating horizontally have a higher chance \referee{of} being scattered and polarization is dominated by those scattered `sideways' at about 90\degr\ having the electric vector perpendicular to the scattering plane, i.e., in the vertical direction.   
The first BH X-ray binary to be observed \referee{by IXPE} in the hard state was \mbox{Cyg~X-1} \citep{Krawczynski2022}. The 2--8 keV polarization was found to align with the radio jet \citep{Miller-Jones2021}, indicating that the corona is extended radially in the disk plane during the hard state. Thus, if the sharp increase of the soft lag magnitude during the HIMS \referee{were} caused by the corona switching from \referee{being} radially extended in the disk plane (scale height $H/R < 1$) to \referee{being} vertically extended along the jet axis ($H/R \gg 1$), we would expect the polarization angle (PA) to flip during the HIMS. However, no such flip was observed when \mbox{Cyg~X-1} was later observed in the soft state \citep[Steiner et al., in prep.]{Dovciak2023}. The coronal emission, which was still strong in the IXPE band, remained aligned with the radio jet. However, the source was not observed in the HIMS during a transition, it was only observed first in the hard state then later in the soft state. Moreover, \mbox{Cyg~X-1} is a persistent source that does not exhibit the usual outburst behaviors such as a q-shaped HID and prominent QPOs. \mbox{4U 1630-47} has also been observed in two states, but these were the soft state \citep{Rawat2023,Ratheesh2024} and the `steep power law' or `very high' state \citep{RodriguezCavero2023}, which is observed reasonably rarely \citep{Remillard2006}. The first \referee{BH X-ray binary} that has enabled IXPE to track the X-ray polarization across a state transition is \source.



\source was discovered on 2023 Aug 24 \citep{GCN.34540,GCN.34544} when it entered a very bright outburst initially reaching $\sim$7~Crab in the 2--20~keV band. All of the source properties strongly point to its identification as a BH X-ray binary, including its X-ray spectrum \citep{ATel16210,ATel16217}, the detection of Type-C QPOs \citep{ATel16215,ATel16219,ATel16247,Mereminskiy2023}, and of \referee{bright} flat-spectrum radio emission indicative of a compact jet \citep{ATel16211,ATel16228}. IXPE first observed \source on 2023 Sept 7 and measured a polarization degree (PD) of $4.1\%\pm0.2\%$ and a PA of $2\fdg2\pm1\fdg3$ \citep{Veledina2023a}, which is roughly in the North-South direction. This aligns with the sub-mm polarization \citep[PA=$-4\fdg1\pm3\fdg5$,][]{ATel16230}, and the optical polarization \citep{ATel16245}. Resolved jet images are not yet available, but using the sub-mm polarization as a proxy suggests that the 2--8 keV polarization again aligns with the jet.

In this paper, we present four further IXPE observations of \source \referee{taken} during a hard to soft state transition and place them in the context of complimentary monitoring by the Neutron star Interior Composition ExploreR (NICER; \citealt{Gendreau2016}) and several radio facilities. We detail our observations and data reduction procedures in Section~\ref{sec:data}, present X-ray polarization results in Section~\ref{sec:results}, and compare with X-ray timing and radio properties in Section~\ref{sec:timing}. We then discuss our results and conclude in Sections~\ref{sec:discussion} and \ref{sec:conclusions}, respectively.

\section{Observations and Data Reduction}
\label{sec:data}

We consider five observations of \source between 2023 September 7 and 2023 October 10 (Table~\ref{tab:obs}) in the context of coverage by NICER and several radio facilities. Figures~\ref{fig:maxi}(a,b) show the timing of the IXPE observations with respect to the source flux variations and the hardness ratio as measured by MAXI \citep{Matsuoka2009}.\footnote{\url{http://maxi.riken.jp/}}  
Figure~\ref{fig:maxi}(c) places the observations on an HID, indicating that they cover the hard to soft state transition.

\begin{table}
\caption{IXPE observation log. \referee{All observations were taken in 2023.}}
\begin{center}
\begin{tabular}{ l l c c c }
\hline
\hline
 Obs & OBSID & Start & End & Live time  \\
     &       & UTC   & UTC & ks  \\
\hline
 1 & 02250901 & Sept 07 19:35 & Sept 08 06:36 & 19.0  \\
 2 & 02251001 & Sept 16 17:15 & Sept 17 13:29 & 37.0  \\
 3 & 02251101 & Sept 27 22:15 & Sept 28 09:51 & 21.0  \\
 4 & 02251201 & Oct 04 12:57 & Oct 04 23:58 & 17.5  \\
 5 & 02251301 & Oct 10 11:30 & Oct 10 22:39 & 17.8 \\
\hline
\end{tabular} 
\end{center} 
\label{tab:obs}
\end{table}

\subsection{IXPE Data Reduction}

IXPE is the first dedicated X-ray polarimetric mission, which measures the polarization in the 2--8 keV band \citep{Weisskopf2022}. 
It carries three X-ray telescopes, each made of a Mirror Module Assembly \citep{Ramsey2022} and a polarization-sensitive gas-pixel detector unit  \citep{Soffitta2021,Baldini2021}, that enable imaging X-ray polarimetry of extended sources and a huge increase of sensitivity for point-like sources.
IXPE has an angular resolution of $\lesssim 30\arcsec$ (half-power diameter, averaged over the three detector units).
The overlap of the fields of view of the three detector units is circular with a diameter of 9\arcmin; spectral resolution is better than 20\% at 6~keV.

We downloaded \referee{Level} 2 data from the IXPE archive at the HEASARC\footnote{\url{https://heasarc.gsfc.nasa.gov/docs/ixpe/archive/}} and analyzed it with \textsc{ixpeobssim} version 30.6.2 \citep{Baldini2022} and HEASOFT/\textsc{xspec} version 12.13.1d \citep{Arnaud1996}. All IXPE observations were carried out with the `gray' filter in front of the detector units \citep[\referee{DUs;}][]{Ferrazzoli2020,Soffitta2021}. \referee{The gray filter} reduces the incident count rate by a factor of $\sim 10$ to reduce the effects of dead-time (which is $\sim 1.2$ ms) and to enable all of the data to be transmitted within the allocated telemetry. Despite its name, the filter is dependent on photon energy, with transmission dropping sharply at low energies. Specific response matrices therefore must be used for observations utilizing the gray filter, which are available both in the \textsc{ixpeobssim} package and in the HEASARC CALDB.

Following \cite{Veledina2023a}, we measure polarization only by fitting weighted $I$, $Q$, and $U$ spectra\footnote{\referee{Since IXPE is not sensitive to circular polarization, it is not possible to extract a Stokes $V$ spectrum.}} (created using the \texttt{pha} algorithm of \textsc{ixpeobssim}) with simple phenomenological models in \textsc{xspec}, and not using the \texttt{pcube} algorithm of \textsc{ixpeobssim}. This method properly takes into account the energy response of the instrument, which is important due to the steep drop in filter transmission at low energies. We fit spectro-polarimetric models separately to the three DUs, and combine them for plotting purposes only.

For polarimetry, we extract events from a circular region with 80\arcsec\ radius centered on the source. \referee{This region is chosen to maximize source counts whilst avoiding subtle dependencies of spectral calibration on extraction radius \citep{DiMarco2023}.} No background subtraction is necessary for sources as bright as \source, since in this case the IXPE background is dominated by scattered source emission, even at large offset \citep[hence, the instrumental background itself is negligible and can be ignored,][]{DiMarco2023}. For timing analysis, we instead use a 180\arcsec\ radius to maximize counts \referee{whilst excluding} events from the detector edges.

\subsection{NICER data reduction}

We analyze four sets of NICER observations of \source that were performed \referee{within 24 hours of}
IXPE observations 1, 2, and 4 (\source was not visible to NICER during IXPE observations 3 and 5). NICER is an X-ray telescope with excellent timing capabilities located on the International Space Station (ISS), and it is sensitive to X-rays in the 0.2--12 keV range. The ObsIDs of the data used are 620980104 on August 28, 6750010501 and 6750010502 on September 7 and 8, 6557020201 and 6557020202 on September 16 and 17, and 6557020401 on October 4. We downloaded these data from the NICER archive on the HEASARC website.\footnote{\url{https://heasarc.gsfc.nasa.gov/docs/nicer/archive/}}
NICER detects X-rays with individual detectors and consists of 7 Modular Power Units (MPUs), each consisting of 8 Focal Plain Modules (FPMs), for a total of 56 detectors, of which 52 have been functional after its launch. The very high source count rates from \source can cause internal telemetry saturation when all FPMs are turned on, resulting in many very short ‘shredded’ Good Time Intervals (GTIs) and preventing reliable timing analysis.\footnote{\url{https://heasarc.gsfc.nasa.gov/docs/nicer/analysis\_threads/heasoft632/}} To mitigate this effect, a reduced number of 17 FPMs were active on August 28 and October 4, 12 FPMs on September 7 and 8, and 10 FPMs on September 16 and 17, which resulted in several hundred seconds of usable data for each ISS orbit. In some orbits, the number of FPMs was reduced even further. We excluded those orbits from our analysis.

NICER has suffered from a light leak problem since 2023 May 22, which increases background noise for observations taken during orbit day, when solar light reaches the X-ray detectors. Detector resets known as undershoots, which are a regular feature of a functioning FPM, happen much more often due to the solar light and can then cause degraded spectral resolution and increased noise levels at low energies.\footnote{\url{https://heasarc.gsfc.nasa.gov/docs/nicer/analysis\_threads/undershoot-intro/}} 
The data were reprocessed with the \texttt{nicerl2} pipeline from NASA's HEASoft \citep[\referee{v6.32.1;}][]{FTOOLS_2014}, released on 2023 August 23, with the recommended and most up-to-date filter columns settings. We used the default \referee{maximum} undershoot rate criterion of 500 \referee{undershoots per second}
for all ObsIDs except 6557020201 and 6557020202, for which we manually set that threshold to 800 \referee{undershoots per second}.
\referee{We accept this higher than usual rate so as to limit the amount of data we need to throw away, but} note that doing so increases the systematic errors on any spectral or timing results from these data, and take extra care in interpreting our results. Otherwise, we followed the recommendations on the HEASARC website. For parts of the NICER timing analysis, we make use of the \textsc{stingray} libraries \citep{Huppenkothen2019a,Huppenkothen2019b,matteo_bachetti_2023_7970570}.

\begin{figure*}
\centering 
\includegraphics[width=2.1\columnwidth,trim=0.0cm 0.0cm 0.0cm 0.0cm,clip=true]{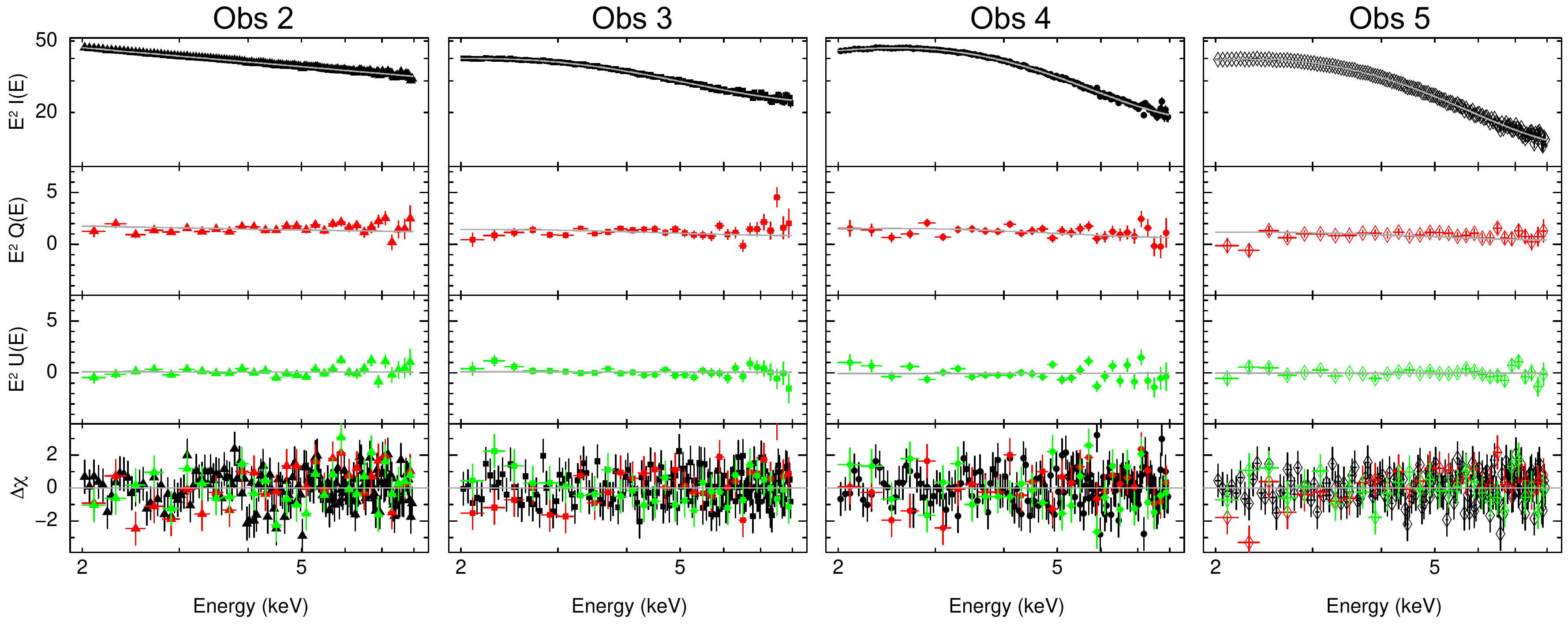}
\caption{Spectro-polarimetric fits for IXPE observations 2--5 (a similar plot for observation 1 can be found in Figure 2 of \citealt{Veledina2023a}). The top three panels show unfolded IXPE Stokes $I$ (black), $Q$ (red), and $U$ (green) spectra (as labeled), in units of keV\,cm$^{-2}$\,s$^{-1}$, along with the best-fit \texttt{polconst} model in gray (\texttt{polconst*powerlaw} for observations 1--3 and \texttt{polconst*[diskbb+powerlaw]} for observations 4--5). The bottom panel shows the residuals in terms of contributions to the fit statistic $\chi$. The three IXPE DUs are grouped together for plotting purposes, but are treated separately in the fit. Data are also re-binned in energy for plotting purposes only (\textsc{xspec} command \texttt{setplot rebin 10.0 5}).}
\label{fig:ixpe_spec}
\end{figure*}

\subsection{Radio Data Reduction}

The RATAN-600 radio telescope observed daily at 4.7, 8.2 and 11.2 GHz. Observations were performed using the `Southern sector and Flat mirror' antenna from  2023 September 14 up to September 30, then with the `Northern sector' (which has a higher sensitivity) from October 1 until October 20. Additional observations at 22.3 GHz were carried out during radio flares. \referee{We calibrated} the data against the quasar 3C~161 (J0627$-$05), included  in the radio flux scale by \citet{Ott1994}. \referee{We estimate the systematic uncertainty to be less than 3\%, which does not affect uncertainties in the spectral index}

The Medicina 32-m radio telescope performed observations (quasi-)simultaneously with the IXPE sessions on 2023 September 15, 28, and October 3 
at 8.4 and 24.2 GHz, depending on the weather and radio interference conditions. We also carried out additional sessions to better follow the evolution of the radio flux density. We applied gain curve and pointing offset corrections to the measurements. The data were calibrated by performing cross-scans on NGC~7027 and 3C~48.
\referee{We estimate uncertainties, including statistical (e.g., from standard deviation of different calibration measurements) and systematic (e.g., from setting the flux density scale), to be 5\%.}
\referee{We performed} the data analysis with the \textsc{Single-Dish-Imager} \citep[\textsc{sdi};][]{Egron2017}, a software designed to perform automated baseline subtraction, radio interference rejection and calibration. 

The Australia Telescope Compact Array (ATCA) observed at 5.5 and 9\,GHz on 2023 September 28, coinciding with 
a portion of IXPE observation~3. 
These observations were taken under project code C2601 as part of a large radio monitoring campaign of \source. ATCA was in a compact H168 configuration,\footnote{\url{https://www.narrabri.atnf.csiro.au/operations/array_configurations/configurations.html}} consisting of 5 antennas in a compact core, and a single, isolated antenna 6\,km away.
We used PKS~B1934$-$638 and PKS~B1730$-$130 for primary and secondary calibration, respectively. The data were analyzed and imaged using standard procedures within the Common Astronomy Software Applications for Radio Astronomy (\textsc{casa}, version 5.1.3; \citealt{2022PASP..134k4501C}). \referee{We included conservative systematic uncertainties on the absolute flux density scale of 4\% \citep[e.g.,][]{2010MNRAS.402.2403M}} by adding in quadrature to statistical errors on the flux density measurement. The full radio campaign will be presented in a future paper.

The 100-m telescope at Effelsberg observed on 2023 September 28, October 10, and October 15 at 5.5 and 14.4 GHz. \referee{We corrected our measurements}
for atmospheric attenuation and the gain-elevation effect. The final calibration in Jy was done with the flux density calibrators 3C~286 and NGC~7027. Details of the analysis procedure are described e.g., in \citet{Kraus2003}

\section{X-ray Polarization Results}
\label{sec:results}

We measure the PD and PA by fitting phenomenological models in \textsc{xspec} to the Stokes $I$, $Q$, and $U$ spectra of all \referee{five} observations. We first use the multiplicative model \texttt{polconst} to impart a constant PD and PA to the entire Stokes $I$ spectral model. In line with what \referee{was} found for observation 1 \citep{Veledina2023a}, we find that we cannot constrain line of sight absorption models for any of the observations, with the best-fit hydrogen column density always tending to zero. Also in agreement with \citet{Veledina2023a}, we find that the \texttt{gain fit} functionality of \textsc{xspec} is required to circumvent calibration issues. The full model we fit to all of the observations is
\begin{equation}
\texttt{polconst*constant*(diskbb+powerlaw)}.
\end{equation}
Here, the constant accounts for the different absolute flux calibration of the different DUs. We freeze it to unity for DU1 and leave it as \referee{separate free parameters} for DU2 and DU3. \texttt{diskbb} is a multi-temperature disk blackbody with peak temperature $kT_{\rm in}$ and normalization \texttt{norm}$_{\rm d}$, and \texttt{powerlaw} models the specific photon flux as \texttt{norm}$_{\rm po}\times E^{-\Gamma}$ photons cm$^{-2}$ s$^{-1}$ keV$^{-1}$.
\begin{deluxetable*}{lcccccc}
\tablecaption{Results of polarimetric and timing analyses of the five IXPE observations. 
\label{tab:parameters}}
\tablewidth{0pt}
\tablehead{
  \colhead{Parameter} & \colhead{Units} & \colhead{Obs 1} & \colhead{Obs 2} &
\colhead{Obs 3} & \colhead{Obs 4} & \colhead{Obs 5}    
}
\startdata
\multicolumn{7}{c}{\texttt{polconst * (diskbb+powerlaw)} }\\
\hline
PD & \% & $4.1 \pm 0.2$ & $3.9 \pm 0.1$ & $3.6 \pm 0.2$ & $3.3 \pm 0.2$ & $2.9 \pm 0.2$ \\
PA & deg  & $2.2 \pm 1.3$ &  $1.8 \pm 1.1$ & $2.1 \pm 1.6$ & $-1.1 \pm 1.8$ & $-0.5 \pm 2.3$  \\
$\chi^2$/dof & &  1282/1329 &  1340/1329 & \referee{1239/1327} & 1327/1327 & 1267/1327  \\
\hline
\multicolumn{7}{c}{\texttt{pollin * (diskbb+powerlaw)} }\\
\hline
PD$_5$ & \% & $4.3 \pm 0.2$ & $4.3 \pm 0.2$ & $4.1 \pm 0.3$ & $3.8 \pm 0.3$ &  $3.8 \pm 0.3$ \\
PD$_{\rm slope}$ & \% & $0.3 \pm 0.2$ & $0.5 \pm 0.1$ & $0.6 \pm 0.2$ & $0.5 \pm 0.2$ & $0.9 \pm 0.2$ \\
PA & deg & $2.0 \pm 1.3$ & $1.9 \pm 1.1$ & $1.6 \pm 1.6$ & $-1.4 \pm 1.8$ & $-0.6 \pm 2.2$ \\
$\chi^2$/dof & &  1277/1328 &  1321/1328 & \referee{1229/1326} & 1321/1326 & 1252/1326 \\
F-test & $\sigma$ & 2.2  &  4.4 & \referee{3.3} &   2.5  & 4.0 \\
\hline
\multicolumn{7}{c}{Timing}\\
\hline
$\nu_{\rm qpo}$  & Hz  & $1.349 \pm 0.003$ & $2.788 \pm 0.008$ & $4.16 \pm 0.04$ & $6.7_{-0.2}^{+0.1}$ & $8.0 \pm 0.3$ \\
HWHM$_{\rm qpo}$ & Hz & $0.126\pm 0.004$ & $0.28 \pm 0.01$ & $0.52\pm 0.06$ & $0.38 \pm 0.37$ & $0.43^{+0.35}_{-0.38}$ \\
$\nu_{\rm 2qpo}$ & Hz  & $2.66 \pm 0.026$ & $5.55 \pm 0.07$ & $\equiv 2\nu_{\rm qpo}$ & $\equiv 2\nu_{\rm qpo}$ & $\cdots$ \\
HWHM$_{\rm 2qpo}$& Hz & $0.37 \pm 0.09$ & $\equiv 2{\rm HWHM}_{\rm qpo}$ & $\equiv 2{\rm HWHM}_{\rm qpo}$ & $\equiv 2{\rm HWHM}_{\rm qpo}$ & $\cdots$ \\
$\chi^2/$dof     &      & 101/84          & 127/88          & 112/91          & 38/50 & 41/49 \\
hue$_{(2-8 \rm keV)}$ & deg & $137.6 \pm 1.7$ & $179.8 \pm 2.1$ & $185.7 \pm 3.3$ & $201.5 \pm 7.6$ & $209 \pm 11$   \\
hue$_{(5-8 \rm keV)}$ & deg & $147.6 \pm 6.9$ & $180.8 \pm 9.3$ & $185.6 \pm 15$  & $\cdots$ & $\cdots$ \\
\enddata
\tablecomments{Spectral and calibration parameters used for the spectro-polarimetric fits are presented in Appendix \ref{sec:cal}. We quote the reduced $\chi^2$ of the two spectro-polarimetric models, and the results of an F-test comparison between the two. For the timing analysis, we quote the centroid and width of the QPO fundamental and second harmonic, the reduced $\chi^2$ values of the multi-Lorentzian fits used to determine them, and the power spectral hue measured in two energy bands (2--8 keV and 5--8 keV). For observations 4 and 5, the power spectral hue is unconstrained in the 5--8 keV band.}
\end{deluxetable*}

We find that only a power law is required for observations 1--\referee{2 (including \texttt{diskbb} improves the fit by $\Delta\chi^2 < 0.5$ for two extra free parameters)}, whereas both disk and power-law contributions are required for observations \referee{3}--5 \referee{($\Delta\chi^2=$ 67, 832 and 772 for observations 3, 4 and 5, respectively, again for two extra free parameters)}. We note that this is an oversimplified model, with the true source spectrum very likely also containing reflection features, but we are often unable to detect these with IXPE given its limited band pass. We also note that the simplicity of the model does not affect the polarization measurement, which is of primary interest here. Since we are only trying to measure the PD and PA of the combined model, and are not attempting to separate into components each with its own polarization, all that matters is the goodness of fit. \referee{For example, including \texttt{diskbb} for observations 1 and 2 yields measurements of PD and PA and their uncertainties that are identical to what we present in Table \ref{tab:parameters} within our chosen rounding convention}. We will leave a joint spectral fit considering multiple observatories to a later paper.

\begin{figure}
\centering 
\includegraphics[width=\columnwidth,trim=1.5cm 1.5cm 2.0cm 1cm,clip=true]{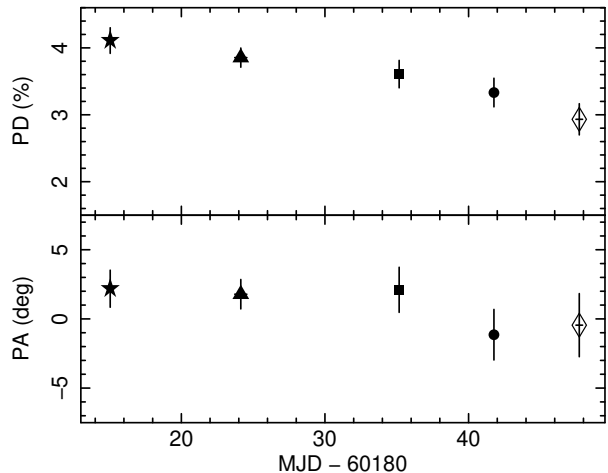}
\caption{Variation of the PD (top) and PA (bottom) in the 2--8 keV range for the five IXPE observations.}
\label{fig:pdpa_mjd}
\end{figure}

\begin{figure*}
\centering 
\includegraphics[width=\columnwidth,trim=1.5cm 1.5cm 2.0cm 10cm,clip=true]{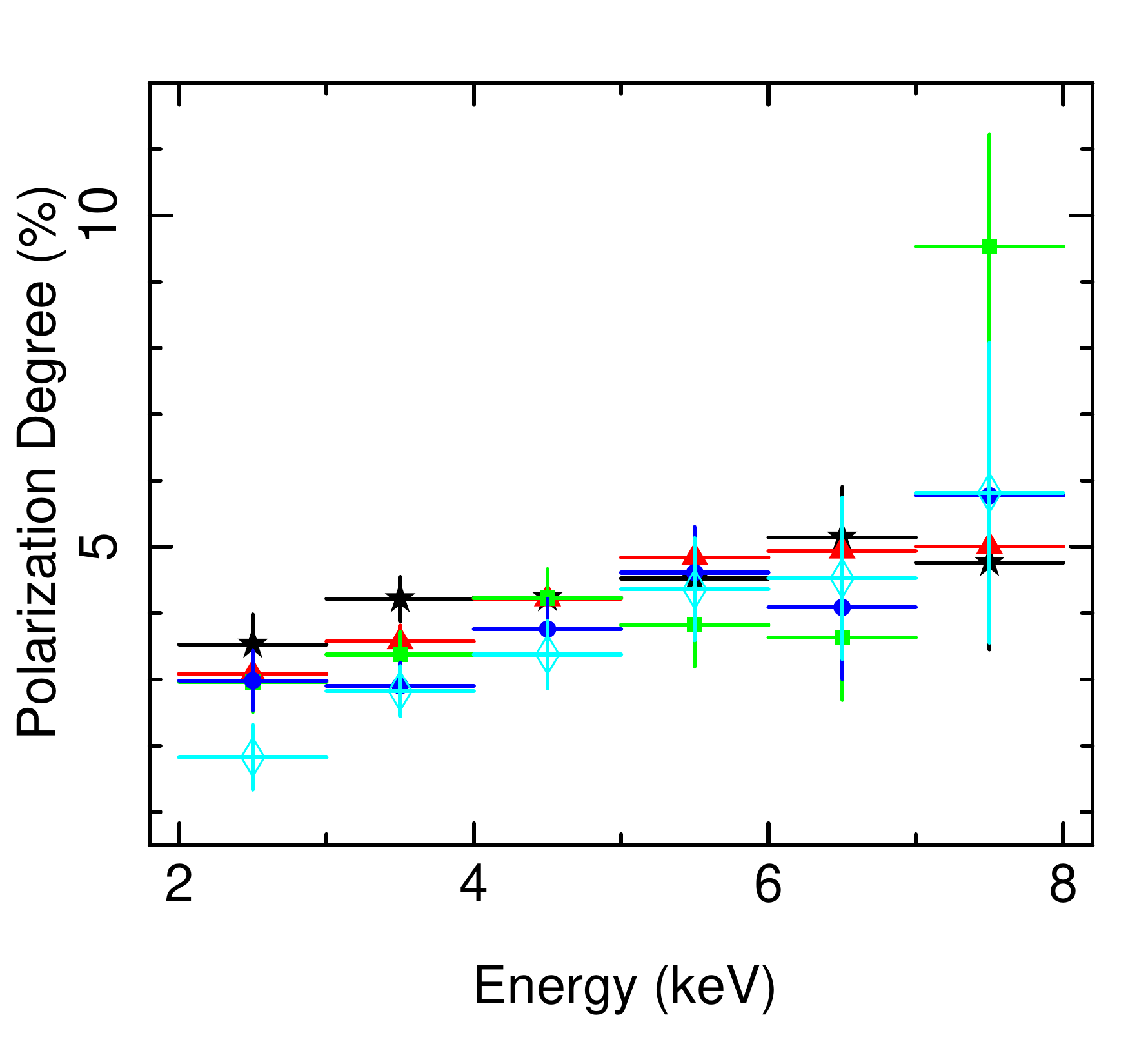}
\includegraphics[width=\columnwidth,trim=1.5cm 1.5cm 2.0cm 10cm,clip=true]{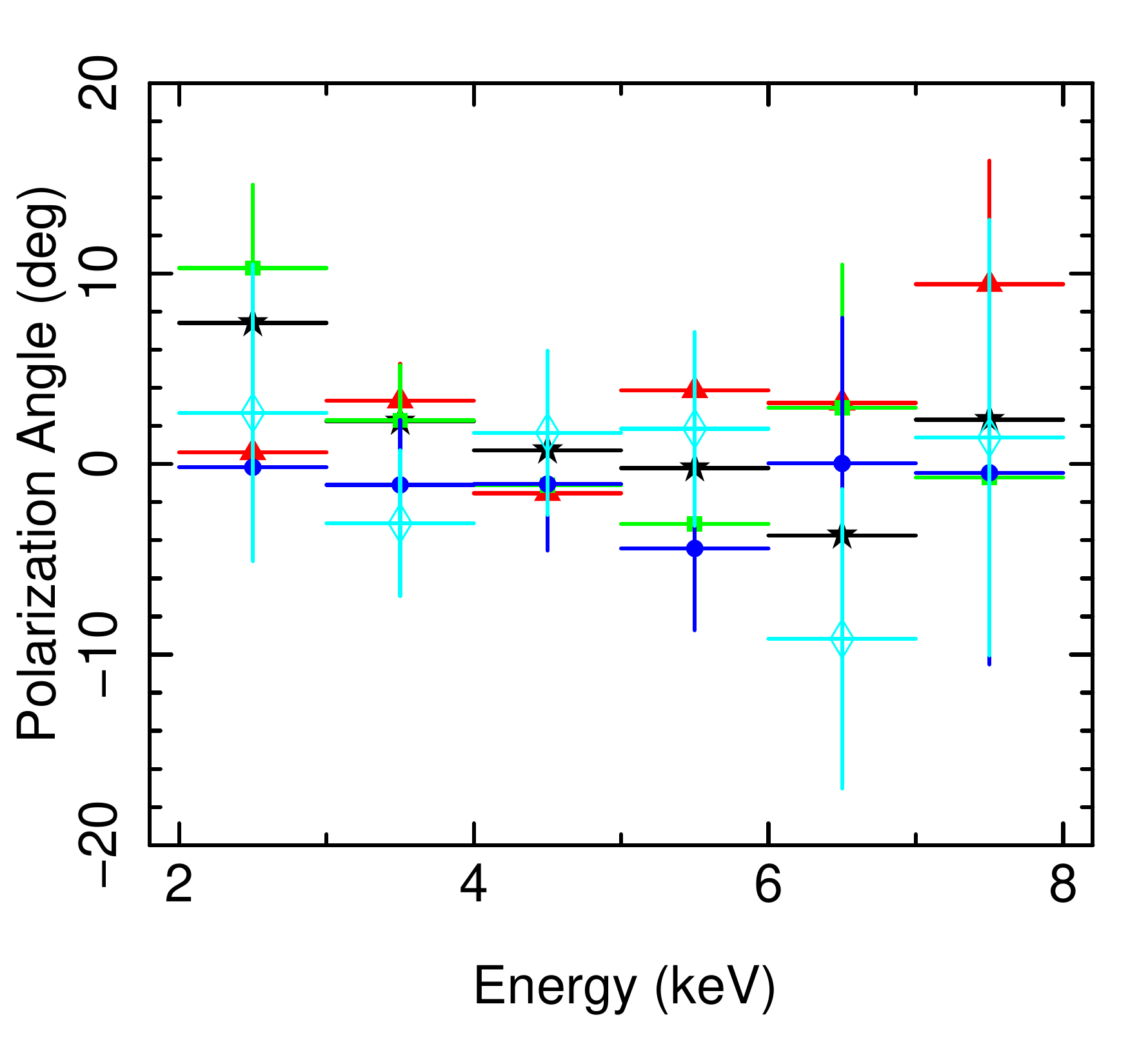}
\caption{PD (left) and PA (right) as a function of energy for observations 1 (black stars), 2 (red triangles), 3 (green squares), 4 (blue circles) and 5 (cyan diamonds). Error bars are 68\% confidence.}
\label{fig:pdpa}
\end{figure*}

We find good fits for all five observations. Figure~\ref{fig:ixpe_spec} shows the $I$, $Q$, and $U$ spectra alongside the best-fit model. The bottom panel demonstrates that there are no structured residuals. The spectral parameters of the phenomenological model used, plus the calibration parameters used, are quoted in Appendix~\ref{sec:cal}. Table~\ref{tab:parameters} lists the best-fit PD and PA alongside the corresponding reduced $\chi^2$ values. The 2--8 keV polarization is detected with high significance for all observations. As can be seen most clearly in Figure~\ref{fig:pdpa_mjd}, the PA is consistent with remaining unchanged across all observations but the PD appears to systematically decrease with time.

We test if the PD is statistically required to change with time by comparing the sum of the best-fit \texttt{polconst} models presented in Table \ref{tab:parameters} ($\chi^2$/dof = \referee{6456/6639}) to alternatives with PD and/or PA tied to be the same across all five observations. We find that holding PA constant in time but allowing PD to vary yields a \referee{fit with a slightly higher null-hypothesis probability}
($\chi^2$/dof = \referee{6459/6643}).
Allowing the PA to change between observations 3 and 4 but stay otherwise constant (motivated by the final two PA values in Figure~\ref{fig:pdpa_mjd} appearing to be $\sim 2\degr$ smaller than the first three) yields no significant improvement to the fit ($\chi^2$/dof = \referee{6456/6642}). We therefore conclude that PA is indeed consistent with being constant in time. Holding both PD and PA constant in time gives $\chi^2/{\rm dof} = \referee{6479/6647}$. Using an F-test, we find that the model with varying PD and constant PA is preferred to the model with constant PD and PA with $3.5 \sigma$ confidence.

To investigate the energy dependence of the polarization, we first freeze all parameters except for PD and PA, and conduct a series of fits that each only consider one of six 1-keV wide energy ranges; i.e., we first fit only in the 2--3 keV range, then only in the 3--4 keV range, and so on. The resulting energy dependent PD and PA values are plotted in Figure~\ref{fig:pdpa}. The PD appears to increase with energy, with the PA consistent with being independent of energy. To test this statistically, we instead fit the model
\begin{equation}
\texttt{pollin*constant*(diskbb+powerlaw)},
\end{equation}
in the 2--8 keV band. The \texttt{pollin} model sets ${\rm PD}(E) = {\rm PD}_1 + {\rm PD}_{\rm slope} (E - 1\,{\rm keV})$ and ${\rm PA}(E) = {\rm PA}_1 + {\rm PA}_{\rm slope} (E - 1\,{\rm keV})$. We find that allowing PA$_{\rm slope}$ to be a free parameter yields no significant improvement in $\chi^2$ over freezing it to zero, thus we fix PA$(E)$ = PA$_1$ = PA hereafter. We replace the constant ${\rm PD_1}$, which is the PD at 1 keV, with another parameter ${\rm PD_5}$, which is the PD at 5 keV. This is so that the constant we fit for corresponds to the PD at an energy within the IXPE bandpass.\footnote{To do this, we define a dummy parameter in \textsc{xspec} that takes the value of PD$_5$.} The two are simply related by ${\rm PD_1} = {\rm PD}_5 - 4~{\rm PD}_{\rm slope}$. The best-fit parameters and goodness of fit of the \texttt{pollin} model are also quoted in Table~\ref{tab:parameters}. The spectral parameters of the two models are identical within our chosen rounding convention. We see that the \texttt{pollin} model is preferred with $>3\sigma$ significance for observations 2, 3, and 5, meaning we can conclude with high statistical confidence that PD increases with energy in these three observations.

We finally test if the slope of PD with energy changes over the five observations. The \texttt{pollin} model with ${\rm PD}_5$, ${\rm PD}_{\rm slope}$ and PA free to vary between observations (as in Table~\ref{tab:parameters}) has $\chi^2/{\rm dof}$ = \referee{6400/6634}. We find that an alternative model with ${\rm PD}_5$ and PA tied across observations gives \referee{an equally good}
fit with $\chi^2/{\rm dof}$ = \referee{6408/6642}. \referee{Given the two fits have the same null-hypothesis probability, we adopt the simpler one as our best fit.} An F-test comparison between this best-fit model and an alternative with all three parameters constant in time ($\chi^2/{\rm dof}$ = \referee{6421/6646}) indicates that the varying slope model is preferred with $2.6\sigma$ confidence. The variation in slope with time is therefore only marginally significant.


\section{Timing Analysis}
\label{sec:timing}

\begin{figure}
\centering
\includegraphics[width=\columnwidth,trim=0.8cm 2.0cm 6.0cm 1.0cm,clip=true]{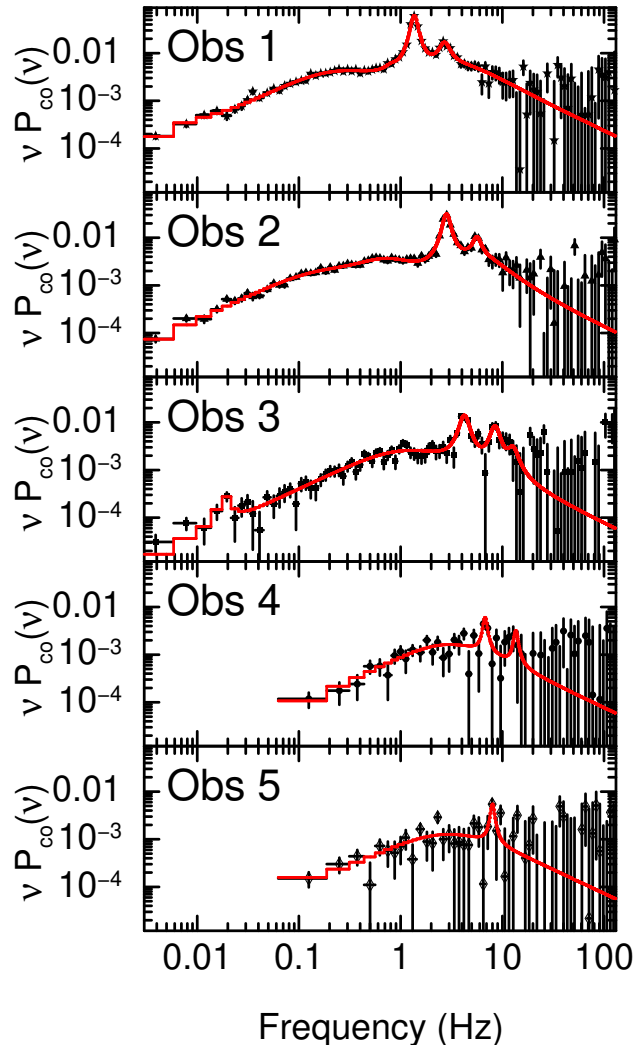}
\caption{Power spectrum of the five IXPE observations, estimated from the co-spectrum of the DU1+DU2 light curve crossed with the DU3 light curve. The best-fit multi-Lorentzian model is shown in red. The 2--8 keV energy band is used for observations 1--3, and the 2--5 keV band for observation 4--5.}
\label{fig:cospec}
\end{figure}


\subsection{IXPE Power Spectrum}

Figure~\ref{fig:cospec} shows a 2--8 keV power spectrum for each of the five IXPE observations. \referee{We calculated these power spectra} by taking the co-spectrum between independent detectors, following \citet{Bachetti2015} and Ewing et al. (in prep.). This method enables us to estimate the power spectrum with no contribution from Poisson noise, thus circumventing the need to model the dead-time-affected Poisson noise contribution. Dead-time is the time after each photon is detected during which another photon cannot be detected. For IXPE this time is $\sim 1.2$ ms, which is relatively large. We first extract two statistically independent light curves with time bins of duration $dt=1/256$ s: one by summing photons from DU1 and DU2, and the other only using DU3 photons. We then calculate the cross-spectrum between the two light curves. To ensure Gaussian distributed uncertainties \citep[see][]{VDK1989,Barret2012}, we ensemble\referee{-}average over segments of length $T_{\rm seg}$ and employ geometric frequency re-binning with re-binning constant $c=1.08$ \citep[\referee{e.g.,}][]{Ingram2012PhDT}. The co-spectrum is the real part of this cross-spectrum.

For observations 1--3, we ensemble\referee{-}average using $T_{\rm seg}=256$ s\referee{, to ensure we have access to a wide range of Fourier frequencies}. Strong, harmonically related peaks can clearly be seen in the resulting co-spectra. These are Type-C QPOs, with centroid frequency increasing during the hard to soft transition in line with known phenomenology \citep{Wijnands1999,Belloni2010,Ingram2019b}. For observations 4--5, we instead use $T_{\rm seg}=8$ s to achieve adequate signal to noise. Using shorter segments leads to less data being discarded due to full segments not fitting into short good time intervals, and the trade-off is losing access to Fourier frequencies below $1/T_{\rm seg}$. The total number of segments averaged over for Obs 1--5 is respectively 67, 117, 70, 2329 and 2297, with the much larger values for the final two observations being due to the much shorter segment length employed (there are 58 and 50 segments of length 256 s in observations 4 and 5 respectively). We still see QPO features consistent with a Type-C classification in these final two observations, but they are only marginally significant
due to far lower signal to noise. Nonetheless, the QPO in observation 4 is confirmed by inspection of the power spectra of the simultaneous NICER observations, in which the QPO features are still very clear. There is no NICER coverage of observation 5 due to Sun constraints.


We model each co-spectrum with a multi-Lorentzian model \citep[e.g.,][]{Psaltis1999,vanstraaten2002}. We use six Lorentzian functions for observation 1, five for observations 2--3, and three for observations 4--5. We decide on the number of Lorentzians \referee{by requiring a component}
to improve the fit \referee{by $\Delta\chi^2>9$ to be included}.
The red lines in Figure~\ref{fig:cospec} represent the best-fit models. The measured QPO frequencies, their half width at half maximum (HWHM) values, and the reduced $\chi^2$ of the fit are listed in Table \ref{tab:parameters}. Subscripts `qpo' and `2qpo' correspond to the first (fundamental) and second harmonic,  \referee{respectively}. For observation 1, the signal to noise is high enough for us to leave the centroid and width of both QPO components free, with the second harmonic consistent with being double that of the first. For other observations, we require the width and/or centroid of the second harmonic to be double that of the fundamental, using an F-test to decide. A third harmonic can be seen in observation 3. We find that the statistically preferred fit is achieved by tying the centroid and FWHM of this component to be triple that of the fundamental. No second harmonic is detected in observation 5.

Our model for observation 3 includes another QPO feature at $18.4 \pm 1.2$~mHz. This feature is marginally significant (\referee{including it improves the fit by $\Delta\chi^2=15$ for three extra free parameters}),
but could be instrumental since IXPE undergoes dithering on three periods: 107, 127, and 900~s (Ewing et al., in prep.). The $\sim 18$~mHz feature could therefore be the second harmonic of the 107~s dithering period. This hypothesis is difficult to definitively test as NICER coverage of observation 3 was not possible (and NuSTAR coverage was also not possible).

\subsection{IXPE Power Colors and Hue}
\label{sec:hue}

\citet{Heil2015a} showed that the evolution of the power spectrum is a more reliable diagnostic of state than the flux spectrum. In analogy to spectral colors, they therefore defined power colors (PCs) as a way to track the shape of the power spectrum with only a few numbers. They defined PC1 as the power spectrum integrated over the 0.25--2 Hz frequency range divided by the integral over the 0.0039--0.03 Hz range. Similarly, they defined PC2 as the 0.031--0.25 Hz integral divided by the 2--16 Hz integral. They showed that, on a plot of PC2 versus PC1, all sources follow the same oval shaped `color wheel'. The evolution around the wheel is clockwise in the outburst rise and anti-clockwise on the decay back to quiescence. Spectral state can therefore be defined purely by the position of the source on this color wheel, which can be parameterized by a single `hue' angle \referee{---} named in direct analogy to a color wheel. The hue is calculated by nominating a center of the color wheel (PC1=4.51920, PC2=0.453724) and then finding the vector that points (in log space) from that center to the (PC1,PC2) point of the observation in question. The hue is then the clockwise angle from a vector in the $(-1,1)$ direction to the vector calculated for the specific observation.

Table \ref{tab:parameters} quotes the hue angle calculated (by following the \citealt{Heil2015} method exactly and employing $T_{\rm seg}=256$ s for all observations) for each observation using the co-spectrum in two different energy bands. The first, hue$_{(2-8)}$, is the full 2--8 keV energy band, which provides the best signal to noise. However, \citet{Heil2015a} based their classification on power spectra of the 2--13 keV energy band of the RXTE proportional counter array (which had an effective area peaking at $\sim$10 keV). To account for the harder band pass of RXTE, \citet{Wang2022} used the 4.8--9.6 keV energy band to unify their analysis of NICER data with the original RXTE analysis of \citet{Heil2015a}. To follow suit, we therefore select the 5--8 keV band pass of IXPE and quote the results in the table as hue$_{(5-8)}$. We see this as the more reliable measurement when signal to noise is sufficiently high. However, the 5--8 keV hue is completely unconstrained for observations 4 and 5 (i.e., the uncertainty estimate is $\pm 180\degr$). We note that the two bands yield similar results for the first three observations, and so tentatively use the 2--8 keV hue for the final two.

\begin{figure}
\centering
\includegraphics[width=\columnwidth,trim=0.0cm 0.0cm 0.0cm 0.0cm,clip=true]{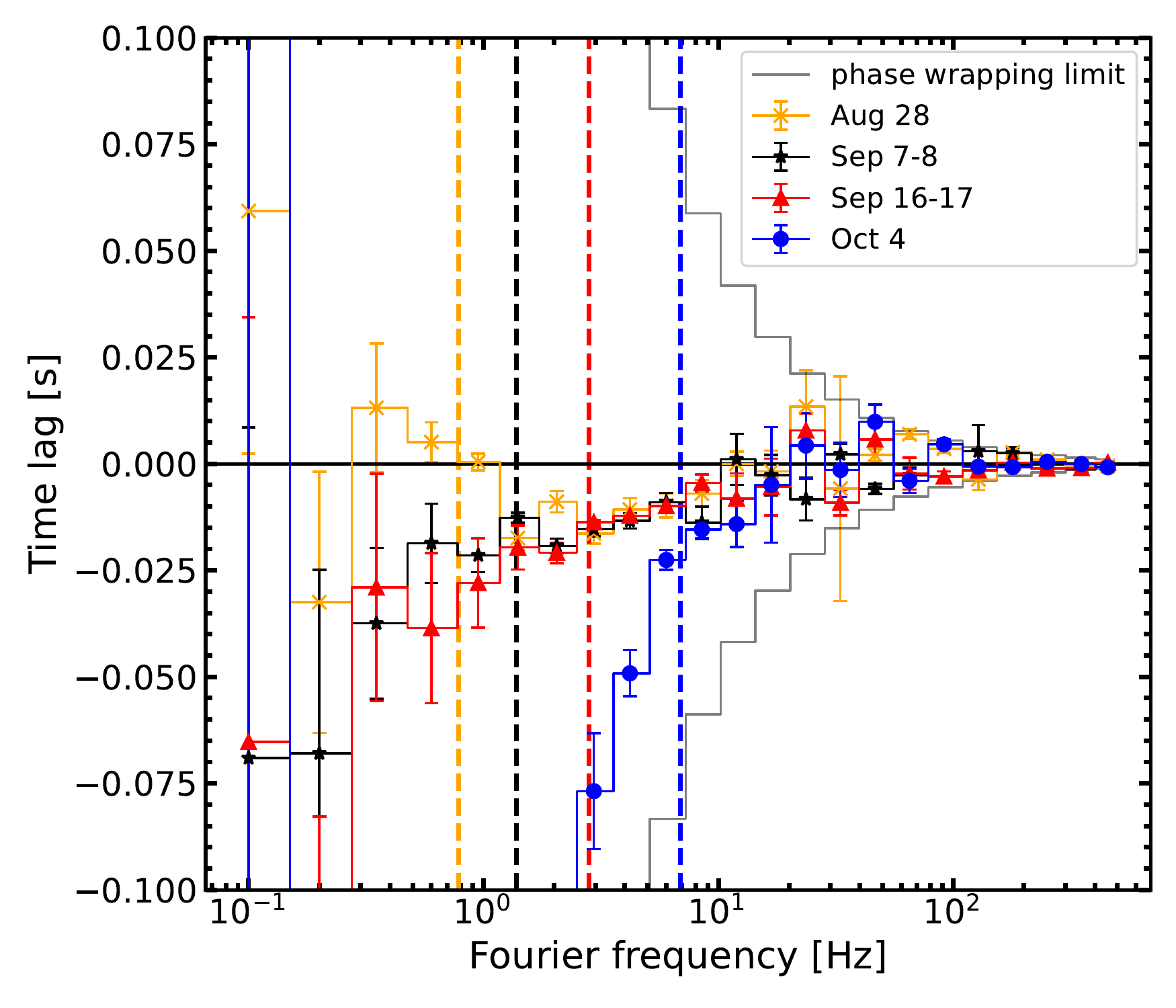}
\caption{Time lag between 2--5  and 0.5--1 keV bands as a function of the Fourier frequency for NICER observations before (Aug~28, orange crosses) and during the first (Sept~7, black stars), second (Sept~16--17, red triangles), and fourth (Oct 4, blue circles) IXPE observations. Positive lag corresponds to hard photons lagging soft photons. The QPO frequencies are marked by the dashed vertical lines. The maximum and minimum measurable time lags (the phase-wrapping limits) are also shown (gray stepped line).}
\label{fig:lagfreq}
\end{figure}

\citet{Heil2015a} defined the hard state as hue$<140\degr$ and the HIMS as $140\degr <$hue $<220\degr$. Observation 1 is therefore on the border between the two states, Obs.\,2--4 are firmly in the HIMS, and observation 5 is consistent within uncertainties with the HIMS or SIMS. Other than a hue $>220\degr$, the onset of the SIMS is also associated with a sharp drop in total rms variability amplitude \citep{Belloni2010}. We do not see such a drop here (the rms is comparable within uncertainty between observations 4 and 5), and so conclude that observation 5 is in the HIMS, at least according to its X-ray timing properties.

\begin{figure}
\centering
\includegraphics[width=\columnwidth,trim=8.5cm 0.0cm 8.5cm 0.0cm,clip=true]{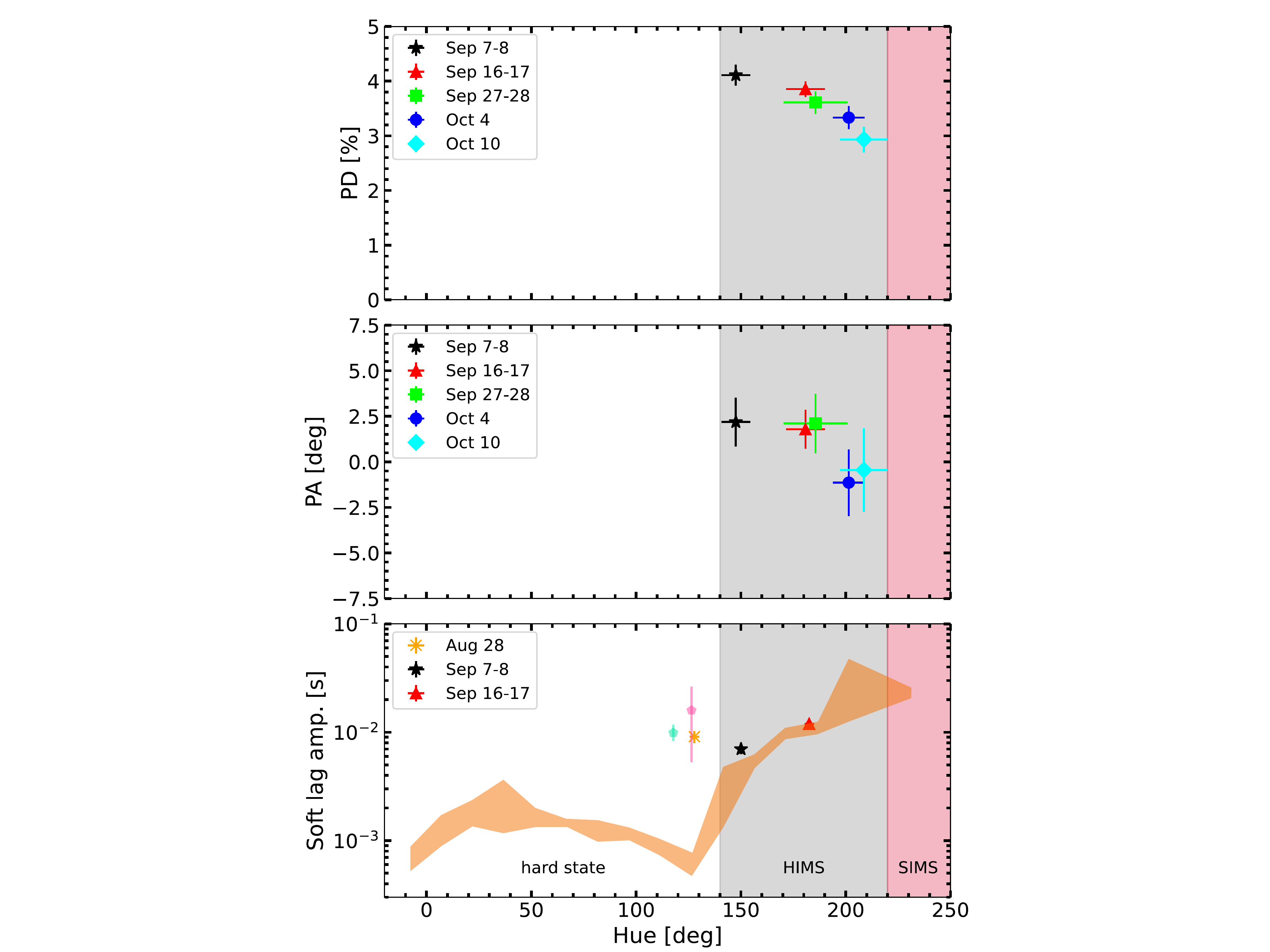}
\caption{PD (top), PA (middle) in the 2--8 keV range, and soft lag amplitude (bottom) as a function of power spectral hue (hue is measured in the 5--8 keV range for Obs 1--3 and in the 2--8 keV range for Obs 4--5). The top two panels are made entirely with IXPE data and the bottom panel entirely with NICER data. The soft lag amplitude is averaged from the time lag versus frequency plots in Figure~\ref{fig:lagfreq}, and the same symbols and colors are employed. The orange shaded area depicts the average over eight sources that \citet{Wang2022} found to follow a common trend. The two outliers to that trend are also plotted: MAXI J1803$-$298 (turquoise pentagon) and EXO 1846$-$031 (pink pentagon).}
\label{fig:laghue}
\end{figure}

\subsection{NICER Time Lags}

Figure~\ref{fig:lagfreq} shows the time lag as a function of Fourier frequency $\nu$ for four NICER observations. \referee{We measure these lags}
using the cross-spectrum \citep{vanderKlis1987} between the 2--5 and 0.5--1 keV light curves. We use a time bin duration of $dt = 0.001$ s, a segment length of $T_{\rm seg}=10$ s, and a re-binning factor of $c=1.4$.
We obtain the time lag by dividing the argument of the cross spectrum by $2\pi\nu$, employing the convention that positive lag corresponds to the harder band lagging the softer band. We also plot the phase-wrapping limits $\pm 1/(2\nu)$ (gray stepped line), which represent the maximum and minimum measurable lags.
The errors on the lags are calculated using the coherence \citep{Uttley2014}. 

We take steps to circumvent the affects of NICER dead-time, which causes spurious anti-correlations between energy bands due to the detector sensitivity becoming anti-correlated with count rate on time scales comparable to or shorter than the dead-time. This effect is usually negligible in NICER due to its small dead-time. However, \source~is exceptionally bright and NICER is currently suffering from solar light leaking in to the detectors, meaning that the spurious anti-correlation dominates on short ($\lesssim0.1$~s) timescales if standard methods are employed. To circumvent this effect, we use one set of detectors to extract the hard band light curve, and another to extract the soft band light curve. We require that no MPUs are used in both light curves because telemetry saturation occurs per MPU and thus introduces spurious anti-correlations between different FPMs within the same MPU. We confirm that this procedure works as expected for the observations plotted in Figure~\ref{fig:lagfreq} by confirming that it returns time lags consistent with zero between full band light curves extracted from these two sets of detectors.

\begin{figure*}
\centering 
\includegraphics[width=1.9\columnwidth,trim=2.0cm 3.5cm 2.0cm 4.0cm,clip=true]{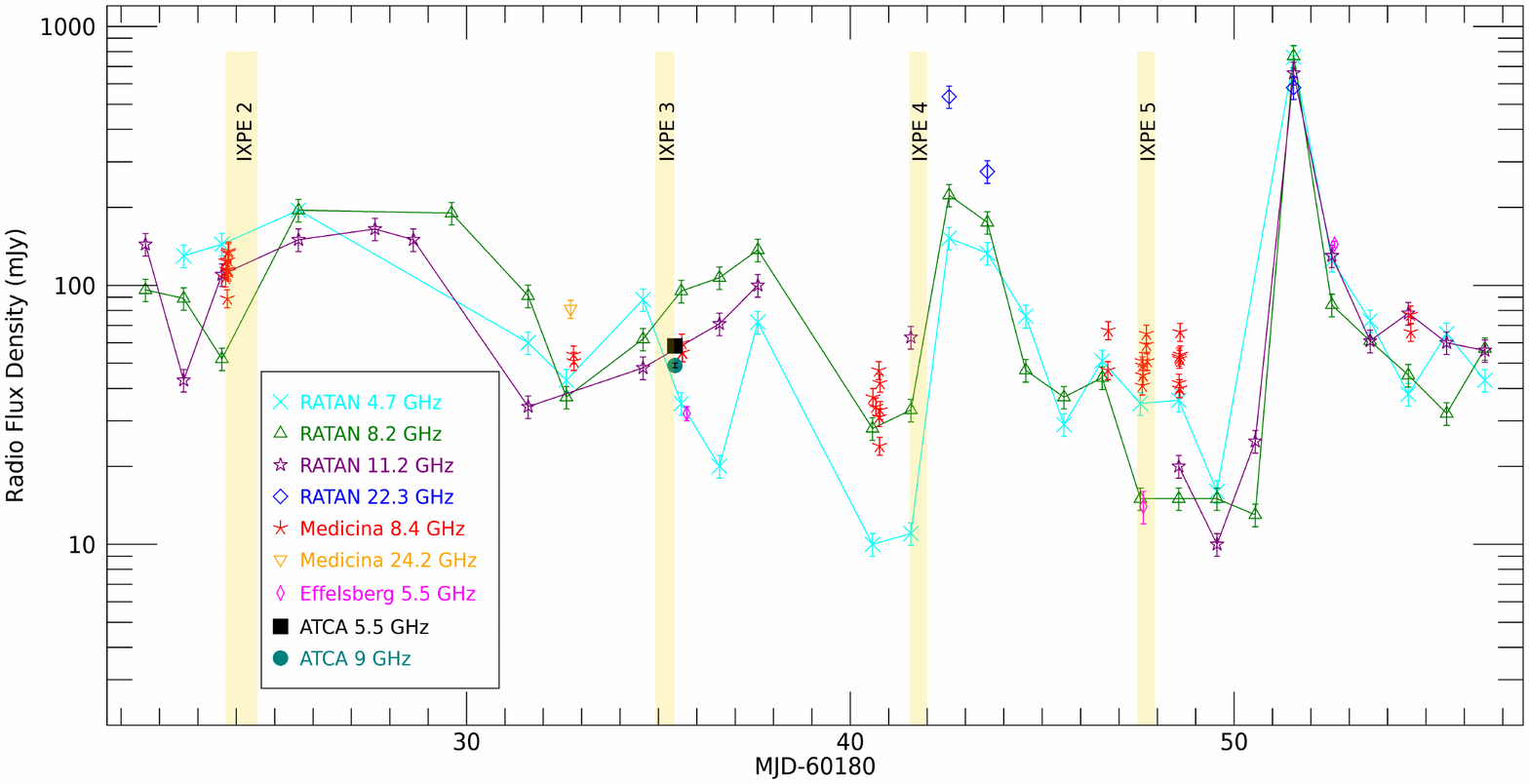}
\caption{Evolution of the radio flux density in the time period encompassing IXPE observations 2--5. Facilities and frequency bands are as labeled.}
\label{fig:radio}
\end{figure*}

We see that soft (negative) lags are visible in Figure~\ref{fig:lagfreq} for a range of Fourier frequencies. To compare these observations to the wider population, we calculate an average soft lag for each, following the exact procedure detailed in \citet{Wang2022}'s population study. This involves averaging the time lag, weighted by the uncertainty, in a frequency range spanning from a minimum $\nu_{\rm min}$ to a maximum $\nu_{\rm max}$, ignoring any frequencies in this range dominated by the QPO (according to a multi-Lorentzian fit to the NICER power spectrum). $\nu_{\rm min}$ is the lowest frequency with a negative time lag. The range $\nu_{\rm min}$ to $\nu_{\rm max}$ is defined as the range of consecutive frequencies in which the phase lag is 1) consistent within $1\sigma$ uncertainty with being negative, and 2) larger (less negative) than $-\pi/2$. The latter stipulation is to avoid the average being biased by phase-wrapping at high frequencies where the lag is at risk of being consistent within uncertainties with the phase-wrapping limit (phase lag of $\pm \pi$).

Figure~\ref{fig:laghue} (bottom) shows the resulting average soft lag amplitude for three of the NICER observations, plotted against 4.8--9.6~keV power spectral hue (see Section \ref{sec:hue}). The two measurements in the HIMS (black star and red triangle) are consistent with the common trend followed by eight of the ten sources in the \citet{Wang2022} sample (orange shaded area). In contrast, the one measurement in the hard state (orange cross) lies far from the common trend, but is remarkably consistent with the two outliers (pentagonal markers) identified by \citet{Wang2022}. These outliers correspond to two sources. One, MAXI J1803$-$298, has only one observation whereas the other, EXO 1846$-$031, has a second observation at hue$\approx 200\degr$ with a large soft lag of $\approx 100$~ms. \referee{As an important caveat, we note that the time lags for these two outlier sources were calculated by stacking separate observations with a slightly greater diversity of properties than the default criteria used in the \citet{Wang2022} methodology and also requiring a smaller signal-to-noise ratio (see their Table 1 and Section 3.2 for details). As such, the soft lags in both sources may be less constrained than shown in Figure~\ref{fig:laghue}. However, we would expect that averaging over observations with different properties would dilute the lag rather than enhance it.}

We do not plot an average soft lag amplitude for the Oct 4 observation. This is because the phase lag of the Oct 4 observation switches from positive in the first frequency bin to $< -\pi/2$ in the second frequency bin; i.e., $\nu_{\rm max}=\nu_{\rm min}$ due to the phase-wrapping limit stipulated in the procedure described above. When we ignore the phase-wrapping limit, we measure an average soft lag amplitude of $35 \pm 3$ ms, which is consistent with the common trend (i.e., orange shaded area). However, we choose not to include the result in Figure~\ref{fig:laghue} because it is calculated using different assumptions to all of the other observations considered. Above the soft lags, we plot 2--8 keV PD (top) and PA (middle) for the five IXPE observations. For observations 1--3, we use the 5--8 keV IXPE hue, which agrees very well with the 4.8--9.6 keV NICER hue for observations 1 and 2. For observations 4--5, we are instead limited to the 2--8 keV IXPE hue. We see that the five IXPE observations span the entire HIMS.

\subsection{Radio Evolution}
\label{sec:radioresults}

Figure~\ref{fig:radio} shows the radio light curve compiled from observations by RATAN-600, ATCA, Medicina and Effelsberg. The radio flux density was highly variable, with the Medicina observations in particular revealing short timescale variability. We see that mini-flares followed IXPE observations 2 and 3. A larger flare followed IXPE observation 4, consistent with the flare detected by the Very Large Array at the same time \citep{ATel16271}. Such a flare often marks the transition to the SIMS, accompanied by the onset of transient ejections, Type-B QPOs and preceding the eventual transition to the soft state \citep{Fender2004,Homan2020}. However, the transition to the SIMS did not occur, and the radio flux stabilized by the time of IXPE observation 5. An even larger radio flare was then observed in the days after observation 5. This flare may be associated with a discrete ejection and the transition to the SIMS. We see in Figure \ref{fig:maxi} that the X-ray spectrum did indeed continue to soften after the final IXPE observation, and is likely in the soft state at the time of writing. However, NICER and IXPE were Sun constrained by this time, precluding detailed determination of the X-ray spectral and timing properties.

We are additionally able to measure polarization at 5.5 and 9\,GHz for the ATCA observation, which was partly simultaneous with IXPE observation 3. We use the un-polarized source PKS~B1934$-$638 to solve antenna leakages (D-terms), and exclude the distant antenna CA06, which is located 6\,km from the relatively compact core of the array. We measure linear polarization with a PD of
$7 \% \pm 2 \%$ and $4\% \pm 2\%$ at 5.5 and 9\,GHz respectively.
No significant circular polarization was detected, with a 3$\sigma$ upper-limit of 0.1\% on Stokes $V$ at both frequencies. With the \textsc{casa} task \texttt{qufromgains}, we determine a PA of $-7\fdg8 \pm 1\fdg5$ at 5.5\,GHz and $-1\fdg7 \pm 0\fdg5$ at 9\,GHz. 
We then determine the PA intrinsic to the source, the electric vector position angle (EVPA), by accounting for Faraday rotation, which is the wavelength-dependent rotation of the polarization vector that occurs as the emission propagates from the source to the observer. The EVPA relates to the PA measured at different wavelengths $\lambda$ as ${\rm EVPA} = {\rm PA}(\lambda) - {\rm RM}\lambda^2$, where RM is rotation measure. We find RM$=-55 \pm 20$\,rad\,m$^{-2}$, yielding an EVPA of $2\degr \pm 2\degr$, consistent with the X-ray PA. \referee{This measurement of RM also confirms that foreground Faraday rotation is negligible at X-ray wavelengths.}

\section{Discussion}
\label{seck:discussion}

We have tracked the X-ray polarization of \source~across a hard to soft state transition with five IXPE observations. This is the first time such an analysis has been possible for a transient X-ray binary. Our observations cover the entire HIMS, with the first observation coinciding with the transition from the hard state and the fifth close to the transition to the SIMS. This is indicated by the track the source followed on the HID (Figure~\ref{fig:maxi}c), the evolution of the power spectrum including QPOs with frequency evolving from $\approx 1.3 - 8$ Hz (Figures~\ref{fig:cospec} and \ref{fig:laghue}), and by the detection of radio flares before and after observation 5 (Figure~\ref{fig:radio}).

Whereas the X-ray PA remains constant throughout the transition, the PD slowly decreases with time (with $3.5\sigma$ confidence, Figure~\ref{fig:pdpa_mjd}). The PD appears to increase with energy in all observations (Figure~\ref{fig:pdpa}), but we can only state this with $>3\sigma$ statistical confidence for observations 2, 3, and 5. The slope of this energy dependence also appears to grow steeper as the spectrum softens, although only with $2.6\sigma$ confidence. The PD at 5 keV is consistent with remaining constant in time. It is therefore possible that the PD of the corona is constant in time, and the overall PD is reducing only because the disk is coming further into the IXPE bandpass as its temperature increases. Such a reduction will happen if the disk is polarized perpendicular to the corona (or un-polarized),
and this scenario would explain the reduction in PD occurring only at lower energies. This picture can be tested using spectro-polarimetric fits involving other observatories such as NICER and NuSTAR, which provide the spectral bandwidth that IXPE lacks. Such a study is beyond the scope of this paper, and we will leave it to future work. If the corona does have a constant PD and PA, this indicates that its aspect ratio $H/R$ remains constant throughout the transition, even if its radial extent is shrinking as in the truncated disk model. Alternatively, or additionally, the reduction in PD could be driven by the corona itself, perhaps from the aspect ratio slightly tending closer to unity (i.e., getting closer to spherical) 
during the transition.


We observe significant evolution in the spectral and timing properties of \source during the HIMS, in line with known phenomenology \citep[e.g.,][]{Belloni2010}. The observed evolution of X-ray binaries is thought to be driven by changes in the geometry of the accretion disk and X-ray corona. There is much uncertainty as to the shape of the corona. The shapes that have been discussed in the literature can be grouped into two main classes: flattened in the disk plane \citep[radially extended;][]{Haardt1991,Esin1997,Poutanen1997}, or confined to a region around the disk axis, along the jet \citep[vertically extended;][]{Kylafis2008,Reig2015,Kara2019,Wang2021}. Some models expect the disk to be truncated in the hard state and to move towards the BH during the transition, whereas other studies infer the disk to already be at the ISCO in the bright hard state \citep[e.g.,][]{Garcia2015,Liu2023a}. The X-ray spectrum alone can be explained in both scenarios, either with a single Comptonized spectrum and a broad iron line dominated by emission from close to the ISCO \citep[e.g.,][]{Parker2014}, or several Comptonized components from a stratified extended medium and a narrower iron line emitted from a truncated disk \citep{Frontera2001cygx1,Ibragimov2005,Shidatsu2011,Zdziarski2021}.

X-ray timing information enables orthogonal constraints. In particular, soft lags (0.5--1 keV lagging 2--5 keV) seen at high Fourier frequencies ($\gtrsim$1~Hz) have been attributed to reverberation. We compare our X-ray polarization measurements with soft lag measurements extracted from simultaneous NICER data (Figure~\ref{fig:laghue}). We find that the soft lags are fairly long ($\gtrsim$10~ms). In the simplest interpretation whereby the lags are driven only by light-crossing delays, this implies a highly vertically extended corona ($H/R \gg 1$) as the only means of there being a large enough physical distance between illuminator and reflector \citep{Wang2021,Lucchini2023}. However, here we find that the X-ray polarization aligns with both the $\sim$230~GHz polarization measured in the hard state \citep[on Sep 3--4;][]{ATel16230}, and with the 7.25~GHz polarization measured quasi-simultaneously with observation 3 (Section \ref{sec:radioresults}). Since the radio polarization often aligns with the jet direction in the hard state \citep[e.g.,][indicating that the jet magnetic field is primarily toroidal]{Corbel2000,Hannikainen2000,Russell2015}, this implies that the corona is instead radially extended.

We see the most likely resolution to this apparent discrepancy between timing and polarization being that the corona is radially extended, but that the soft lags in the HIMS are dominated by effects other than pure light-crossing delays. Such effects have been considered extensively in the literature. For example, interference between variable Comptonized signals emitted primarily from different regions of a stratified corona \citep{Veledina2018,Mahmoud2019,Kawamura2023}, and/or between variable disk and corona signals \citep{Rapisarda2017,Uttley2023}. These \referee{interference} effects have also been used to explain the complex, energy-dependent evolution of the power spectrum \citep[\referee{e.g.,}][]{Veledina16,Rapisarda2016,Rapisarda2017,Chainakun2021}. 

\referee{Another potentially important effect is the time that reflected photons spend scattering in the disk atmosphere before escaping \citep{Salvesen2022}. This additional time lag is negligible for iron line photons \citep{Garcia2013a}, but may be important for soft emergent photons, many of which have experienced many scatterings. The longest scattering delays are for photons that partially thermalize (requiring the most scatterings). Thus the sudden increase in soft lag could coincide with the disk becoming sufficiently hot for the thermalized photons to enter the 0.5--1~keV band.}

Although light-crossing delays must be present at some level, we suggest that they are not the dominant process in the HIMS, at least for soft X-rays in the frequency ranges accessible to current missions. In contrast, light-crossing delays could still potentially be the dominant mechanism, even in the HIMS, in the iron line energy range, and in frequency ranges ($\gtrsim 100$~Hz) only available to future high-throughput space missions \citep{eXTP2019,Ray2019}. Our conclusion remains relatively tentative until it can be confirmed or falsified by the publication of resolved images of the \source~radio jet, since it hinges on the assumption that the intrinsic radio polarization is aligned with the jet.


The long soft lags we observe in the HIMS are representative of what is observed for the general BH X-ray binary population \citep{Wang2022}. However, we also measure a similarly long soft lag in the hard state (Figure~\ref{fig:laghue}).
Since the \source~lags are very similar to those of two outliers in the \citet{Wang2022} population study, our favored explanation is that these outliers and \source~are actually members of a subset of the X-ray binary population that happened to be under-represented in the NICER archive until now. Although there are 10 sources in the \citet{Wang2022} sample, the hard state is only represented by three sources (reflecting the difficulty of quickly slewing to a source at the onset of an outburst), and is dominated by one (MAXI J1820+070, which evolved exceptionally slowly). Therefore, it is plausible that a large subset of X-ray binaries exhibit long soft lags already in the hard state, but that none of them had been observed in the hard state by NICER at the time of the \citet{Wang2022} analysis.

We can only speculate as to the physical property that separates the two subsets of soft lag evolution with hue. Perhaps the difference is orbital inclination angle. There is evidence that inclination influences other timing properties \citep{Motta2015,vandeneijnden2017}. Misalignment between the BH spin and orbital axes, which was suggested to be large in MAXI~J1820+070 \citep{Poutanen2022,Thomas2022}, could also be a factor. NICER monitoring of sources with a variety of orbital and jet parameters is required for a definitive test.

The constancy of PA we observe here rules out any interpretation of the transition that predicts the PA to flip by $90\degr$ during the HIMS. Such a flip would result from the corona switching from radially extended in the hard state ($H/R<1$) to vertically extended ($H/R\gg 1$) by the end of the HIMS \citep[\referee{e.g.,}][]{Krawczynski2022a}. This scenario was previously plausible, since earlier IXPE results had indicated the corona to be radially extended in the hard state, and interpretation of the long observed soft lags as pure light-crossing delays requires a very large vertical corona height in the HIMS. This picture also provided a ready interpretation of the discrete jet ejection typically detected soon after the soft lag increase as the corona itself being fired away from the BH \citep{Wang2021}. Indeed, it is remarkable that the X-ray polarization does not seem to react to the radio flare between observations 4 and 5, which presumably results from an ejection of coronal plasma. \referee{A similar switch in coronal geometry is inferred from fitting the QPO model of \cite{Karpouzas2020} to Type-C QPO lag-energy spectra of several sources. For several transients, a switch from radially extended to vertically extended is inferred to occur during the SIMS \citep[e.g.,][]{Zhang2022,Ma2023}. In contrast, the corona of GRS 1915+105 is inferred to be vertically extended during the $\chi-$state \citep{Mendez2022}, which is analogous to the hard state. The model can be tested in future by fitting it to NICER data of \source and determining whether or not the inferred coronal geometry is compatible with our polarization measurements.}

\section{Conclusions}
\label{sec:conclusions}

We have made the first X-ray polarimetric observations of a BH X-ray binary during a state transition. With five IXPE observations of \source spanning the HIMS, we find that the PD gradually decreases from $\sim$4\% at the start of the HIMS to $\sim$3\% at the end, and the PA remains in the North-South direction throughout. The PA aligns with sub-mm polarization measured $\sim$ 4 days before the first IXPE observation, and also with the 7.25~GHz radio polarization that we measure quasi-simultaneously with the third IXPE observation. Using the radio PA as a proxy for the jet position angle, this implies that the corona remains radially extended (flattened in the plane of the disk) throughout the transition. This conclusion can be definitively tested upon the publication of resolved radio images.

We find that the PA is consistent with being independent of energy, whereas the PD increases with energy (although only with $>3\sigma$ statistical significance for three of the five observations). These properties are very similar to \mbox{Cyg~X-1} in the hard state.

We detect Type-C QPOs with centroid frequency increasing from $\approx 1.3$ Hz in the first observation to $\approx 8$ Hz in the fifth. We also measure long soft lags (0.5--1 keV lagging 2--5 keV photons) with NICER of $\gtrsim$10 ms, which would require a strongly vertically extended corona to be explained purely by light-crossing delays. Our polarization results therefore imply that the soft lags in the HIMS are dominated by processes other than pure light-crossing delays. We also find that the soft lags of \source do not evolve with state as expected from earlier studies, indicating that it belongs to a sub-population that was under-represented in previous samples.

\section*{Acknowledgments}

We thank the IXPE mission operations staff for their effort and dedication in planning the target of opportunity observations analyzed here: Stephanie Ruswick, Jenny Gubner, Kurtis L. Dietz, Darren J. Osborne, Zach Allen Alexander Pichler, Kacie Davis, Sam Lippincott, Alana Martinez, Alex Fix, Lee Reedy, Deb McCabe, Allison Rodenbaugh, Amelia De Herrera-Schnering, and Cole Writer. We thank Phil Uttley for insightful discussions on the NICER timing analysis and  interpretation of our results. We thank Jingyi Wang for help with measuring the average soft lag amplitude.
IXPE is a joint US and Italian mission.  The US contribution is supported by the National Aeronautics and Space Administration (NASA) and led and managed by its Marshall Space Flight Center (MSFC), with industry partner Ball Aerospace (contract NNM15AA18C).  The Italian contribution is supported by the Italian Space Agency (Agenzia Spaziale Italiana, ASI) through contract ASI-OHBI-2022-13-I.0, agreements ASI-INAF-2022-19-HH.0 and ASI-INFN-2017.13-H0, and its Space Science Data Center (SSDC) with agreements ASI-INAF-2022-14-HH.0 and ASI-INFN 2021-43-HH.0, and by the Istituto Nazionale di Astrofisica (INAF) and the Istituto Nazionale di Fisica Nucleare (INFN) in Italy.  This research used data products provided by the IXPE Team (MSFC, SSDC, INAF, and INFN) and distributed with additional software tools by the High-Energy Astrophysics Science Archive Research Center (HEASARC), at NASA Goddard Space Flight Center (GSFC).
This research has made use of the MAXI data provided by RIKEN, JAXA and the MAXI team.
This work is partly based on observations with the Medicina telescope operated by INAF – Istituto di Radioastronomia (Italy). ATCA is part of the Australia Telescope National Facility (https://ror.org/05qajvd42) which is funded by the Australian Government for operation as a National Facility managed by CSIRO. We acknowledge the Gomeroi people as the Traditional Owners of the ATCA observatory site.
The Effelsberg 100-m radio telescope is operated by the Max-Planck-Institut f\"ur Radioastronomie on behalf of the Max-Planck-Society.

A.I. acknowledges support from the Royal Society. 
A.V. thanks the Academy of Finland grant 355672 for support.
M.D., J.S., and V.Kar. thank GACR project 21-06825X for the support and institutional support from RVO:67985815. 
T.D.R. is an INAF Research Fellow.
H.K. acknowledges support by NASA grants 80NSSC22K1291, 80NSSC23K1041, and 80NSSC20K0329.
The French contribution is supported by the French Space Agency (Centre National d'Etude Spatiale, CNES) and by the High Energy National Programme (PNHE) of the Centre National de la Recherche Scientifique (CNRS).
V.K. acknowledges support from the Finnish Cultural Foundation.
\referee{We thank the anonymous referee for detailed and insightful comments.}


%

\vspace{5mm}
\facilities{IXPE, MAXI, NICER, RATAN-600, Medicina, ATCA, Effelsberg} 



\software {\textsc{ixpeobssim} \citep{Baldini2022}, \textsc{xspec} \citep{Arnaud1996}, 
\textsc{stingray} \citep{Huppenkothen2019a,Huppenkothen2019b,matteo_bachetti_2023_7970570}, \textsc{casa} \citep{2022PASP..134k4501C}, \textsc{sdi} \citep{Egron2017}
}


\appendix

\section{Spectral and calibration parameters}
\label{sec:cal}

Table \ref{tab:cal} lists the spectral and calibration parameters used for our spectro-polarimetric fits in Section \ref{sec:results}. The spectral parameters are well constrained, except for the power law in observations 4--5. Here, the uncertainties are large due to trade offs with the \texttt{diskbb} component in the relatively narrow IXPE band pass. Most calibration parameters are consistent within uncertainties across all observations. However, we were unable to find an acceptable fit when tying all of them to be identical for all observations. There are likely two effects combining to cause this. First, the changing flux and spectral shape of the source could have caused subtly different calibration issues for each observation. This could be due to a known charging effect on the detector gain that depends on the source count rate. Second, the model is over-simplified and so to some extent the gain fit will be fitting real features (e.g., the relativistically broadened iron line), which will be changing in shape and strength from one observation to the next. We note that the gain fit (or a similar procedure) is required to simultaneously fit IXPE data with data from other observatories such as NICER or NuSTAR \citep[e.g.,][]{Svoboda2024}, indicating that the need for the gain fit is not purely down to the model being over-simplified.

\begin{table*}
\caption{Spectral (top) and calibration (bottom) parameters used in the \texttt{polconst} spectro-polarimetric fits. Those used for the \texttt{pollin} models are identical within uncertainties. The constant accounts for the different absolute flux calibration of the different DUs. We utilize the \texttt{gain fit} functionality of \textsc{xspec}. This scales the photon energy to $E^\prime = E / {\rm slope} - {\rm offset}$. Thus the original gain scale is recovered for slope=1 and offset=0.}
\begin{center}
\begin{tabular}{ l l c c c c c }
\hline \hline
                    &  & Obs 1 & Obs 2 & Obs 3 & Obs 4 & Obs 5 \\
\hline
$kT_{\rm in}$  & keV &  -  & - & \referee{$0.96 \pm 0.01$}  & $1.029_{-0.006}^{+0.025}$ & $0.993_{-0.005}^{+0.020}$  \\
\texttt{norm}$_{\rm d}$ & &  -   &  - &  \referee{$2800\pm500$} & $4800_{-400}^{+250}$ & $5000\pm200$ \\
$\Gamma$   & &  $1.80_{-0.01}^{+0.02}$  & $2.26 \pm 0.01$ & \referee{$2.09_{-0.25}^{+0.17}$}  & $1.78^{+0.57}_{-0.70}$ & $1.78_{-0.82}^{+0.73}$  \\
\texttt{norm}$_{\rm po}$ & &  $33.7 \pm 0.6 $   &  $59.8 \pm 0.6$  &  \referee{$29_{-14}^{+7}$} & $11^{+27}_{-9}$ & $8_{-6}^{+23}$ \\
\hline
constant  & DU1 &  1  & 1 & 1  & 1 & 1\\
slope     & DU1 &  $0.940 \pm 0.002$  & $0.941 \pm 0.002$ & \referee{$0.934 \pm 0.006$}  & $0.968 \pm 0.008$ & $0.950 \pm 0.014$ \\
offset    & DU1 &  $0.056 \pm 0.005$  & $0.058 \pm 0.004$ & \referee{$0.070 \pm 0.008$}  & $0.051 \pm 0.010$ & $0.054 \pm 0.017$ \\
constant  & DU2 &  $0.978 \pm 0.001$  & $0.979 \pm 0.001$ & \referee{$0.978 \pm 0.001$}  & $0.979 \pm 0.002$ & $0.978 \pm 0.002$ \\
slope     & DU2 &  $0.935 \pm 0.002$  & $0.933 \pm 0.002$ & \referee{$0.940 \pm 0.006$}  & $0.967 \pm 0.007$ & $0.946 \pm 0.014$ \\
offset    & DU2 &  $0.055 \pm 0.005$  & $0.089 \pm 0.004$ & \referee{$0.089 \pm 0.008$}  & $0.059 \pm 0.010$ & $0.063 \pm 0.017$ \\
constant  & DU3 &  $0.921 \pm 0.001$  & $0.921 \pm 0.001$ & \referee{$0.919 \pm 0.001$}  & $0.921 \pm 0.001$ & $0.915 \pm 0.002$ \\
slope     & DU3 &  $0.942 \pm 0.002$  & $0.940 \pm 0.002$ & \referee{$0.954 \pm 0.006$}  & $0.972 \pm 0.007$ & $0.952 \pm 0.001$ \\
offset    & DU3 &  $0.086 \pm 0.005$  & $0.103 \pm 0.004$ & \referee{$0.064 \pm 0.007$}  & $0.082 \pm 0.010$ & $0.066 \pm 0.016$ \\
\hline
\end{tabular} 
\end{center} 
\label{tab:cal}
\end{table*}


\bibliography{biblio}
\bibliographystyle{aasjournal}

\end{document}